\documentclass[aps,longbibliography,superscriptaddress,onecolumn,preprint,tightenlines]{revtex4-1}
\usepackage{graphicx}
\usepackage{epstopdf}
\usepackage{amssymb,amsmath}
\usepackage{relsize}
\usepackage{placeins}
\usepackage{mathtools}
\usepackage{setspace}
\usepackage{enumitem}
\usepackage{lineno}
\usepackage{tikz}
\usepackage{pgffor}
\usepackage{booktabs}   
\usepackage{lineno,hyperref}
\usepackage{comment}
\usepackage{nicefrac}
\usepackage{todonotes}
\usepackage{natbib}

\def\gdot{\dot{\gamma}}
\def\xv{\boldsymbol{x}}

\def\uv{\boldsymbol{u}}

\def\tauv{\boldsymbol{\tau}}

\DeclareMathOperator{\sign}{sgn}

\newcommand\Bn{\mbox{\textit{Bn}}}
\newcommand\Bnlub{\textit{B}^*}

\newlist{steps}{enumerate}{1}
\setlist[steps]{label=\textit{Step~\arabic*~}}

\newcommand{\revb}[1]{{\color{red} #1}}

\begin{document}

\title{On the viscoplastic squeeze flow between two identical infinite circular cylinders}


\author{A.R.~Koblitz}
\affiliation{Department of Physics, Cavendish Laboratory, J J Thomson Avenue, Cambridge, CB3 0HE, UK}

\thanks{A.R.~Koblitz acknowledges financial support from the EPSRC Centre for Doctoral Training in Computational Methods for Materials Science under grant EP/L015552/1}
\thanks{This work was supported by the Schlumberger Gould Research Centre.}
\email{ark44@cam.ac.uk}
\author{S.~Lovett}
\affiliation{Schlumberger Gould Research Centre, High Cross, Madingley Road, Cambridge CB3 0EL, UK}
\author{N.~Nikiforakis}
\affiliation{Department of Physics, Cavendish Laboratory, J J Thomson Avenue, Cambridge, CB3 0HE, UK}

\date{\today}

\begin{abstract}
Direct numerical simulations of closely interacting infinite circular cylinders in a Bingham fluid are presented, and results compared to asymptotic solutions based on lubrication theory in the gap.
Unlike for a Newtonian fluid, the macroscopic flow outside of the gap between the cylinders is shown to have a large effect on the pressure profile within the gap and the resulting lubrication force on the cylinders.
The presented results indicate that the asymptotic lubrication solution can be used to predict the lubrication pressure only if the surrounding viscoplastic matrix is yielded by a macroscopic flow.
This has implications for the use of sub-grid-scale lubrication models in simulations of non-colloidal particulate suspensions in viscoplastic fluids.
\end{abstract}

\maketitle



\section{Introduction}
\label{sec:introduction}

Complex fluids are ubiquitous in natural and industrial processes, from food processing, to lava or debris flows, to oil and gas applications.
The mechanical behaviour of these fluids arises from the microstructure of the fluid, for example
emulsion droplets and clays in drilling muds, or polymer chains in viscoelastic fluids.
 When non-colloidal particles much larger than the fluid microstructure are added, the system can be thought of
as a particulate suspension in a complex (continuum) fluid.
Examples of these types of systems include fresh concrete and debris flows~\citep{Ovarlez2015}. The hydrodynamic
interaction between particles affects the suspension bulk properties and dynamics and is of great interest. In the case of a
Newtonian fluid, analytical solutions exist for slow flow past spheres and cylinders 
\citep{Stimson1926,Umemura1982} and the squeeze flow between them using asymptotic analysis. Viscoplastic fluids,
of interest to this work, are characterised by a discontinuous nonlinear constitutive equation thereby introducing
additional complexities when analytical solutions are sought.

So far, studies on interacting spheres and cylinders in viscoplastic flows have largely focussed on drag and
pressure drop (in the case of flow past arrays) of collinear arrangements, aligned either parallel or
perpendicular to the flow \citep{Liu2003,Horsley2004,Jie2006,Merkak2006,Yu2007,Tokpavi2009,Jossic2009}. Numerical
studies using the Bingham constitutive law have been found to be in good agreement with experimental work using
Carbopol 940 gels, developing drag correlations and stability criteria (with respect to sedimentation)
\citep{Liu2003,Merkak2006,Tokpavi2009,Jossic2009}. Viscoplastic squeeze flow between coaxial cylindrical disks has
been studied analytically for both planar \citep{Muravleva2015} and axisymmetric
\citep{Smyrnaios2001,Muravleva2017} configurations. The configuration of collinearly approaching bodies in a
viscoplastic flow has received only cursory attention in numerical studies, eg \citet{Tokpavi2009,Yu2007}, with
no examination of the interstitial squeeze flow. 

This study therefore examines the two-dimensional squeeze flow between two approaching infinite circular cylinders in a Bingham 
viscoplastic fluid by direct numerical simulation. The configuration studied is such that the gap between the
two cylinders is small ($1\,\%$ of the cylinder radius). We also make use of the asymptotic analysis by~\citet{Balmforth2017a} to compute leading order lubrication solutions for the squeeze flow between two
approaching cylinders in a Bingham fluid. We compare the analytical and numerical solutions and demonstrate that in a 
quasi-unconfined system the squeeze flow is greatly affected by flow 
external to the gap, but that the asymptotic solution may be recovered under certain flow conditions in the wider
domain.
This is contrary to the Newtonian equivalent, and has implications on using the viscoplastic
lubrication force approximation as a sub-grid-scale model in coarse simulation techniques.

The paper is organised as follows. In section~\ref{sec:numerics} we present the problem of interest
and briefly describe the solution strategy employed for the direct numerical simulations, and the lubrication theory calculations.
In section~\ref{sec:results} we present direct numerical simulations of the quasi-unconfined system.
These are compared to simulations of the domain restricted to the gap only, and to the asymptotic solutions from lubrication theory.
These comparisons demonstrate the influence of the wider flow field on the lubrication pressure.
In section~\ref{sec:conclusions} we discuss the results and the implications for sub-grid-scale modelling.

\section{Mathematical formulation and solution}
\label{sec:numerics}

We consider the slow, steady flow of an incompressible viscoplastic fluid around two rigid, infinite circular
cylinders. The fluid has velocity $\hat{\uv}(\hat{\xv})$, pressure $\hat{p}(\hat{\xv})$ and a symmetric total stress tensor
$\hat{\tauv}-\hat{p}\boldsymbol{\delta}$, where variables with a hat are dimensional.
In the absence of inertia, the conservation of mass is

\begin{equation}
    \frac{\partial \hat{u}}{\partial \hat{x}} + \frac{\partial \hat{v}}{\partial \hat{y}} = 0,\quad
\end{equation} 
and the conservation of momentum is
\begin{align}
    \frac{\partial \hat{\tau}_{xx}}{\partial \hat{x}} + \frac{\partial \hat{\tau}_{xy}}{\partial \hat{y}} -
    \frac{\partial \hat{p}}{\partial \hat{x}} &= 0,\\
    \frac{\partial \hat{\tau}_{yx}}{\partial \hat{x}} + \frac{\partial \hat{\tau}_{yy}}{\partial \hat{y}} -
    \frac{\partial \hat{p}}{\partial \hat{y}} &=0.
\end{align}

As a constitutive law we use the Bingham model

\begin{equation}
    \begin{cases}
        \hat{\tau}_{ij} = \left(2\hat{\eta}+\frac{\hat{\tau}_Y}{\hat{\dot{\gamma}}}\right)\hat{\dot{\gamma}}_{ij} &
        \quad\mbox{if $\hat{\tau}>\hat{\tau}_Y$,}\label{EQ: Bingham model} \\
        \hat{\dot{\gamma}}_{ij} = 0 & 
    \quad\mbox{if $\,\hat{\tau}\leq\hat{\tau}_Y$,}
\end{cases}
\end{equation}
where $\hat{\tau}_Y$ and $\hat{\eta}$ are the yield stress and the plastic viscosity of the fluid, respectively,
$\hat{\dot{\gamma}}_{ij}$ is the rate of strain tensor
associated with the velocity field, and
\begin{align}
    \hat{\dot{\gamma}}_{ij}=\frac{1}{2}\left(\frac{\partial \hat{u}_i}{\partial \hat{x}_j} + \frac{\partial
    \hat{u}_j}{\partial \hat{x}_i}\right),\quad
    \hat{\dot{\gamma}}=\sqrt{\frac{1}{2}\hat{\dot{\gamma}}_{ij}\hat{\dot{\gamma}}_{ij}},\quad
    \hat{\tau} = \sqrt{\frac{1}{2}\hat{\tau}_{ij}\hat{\tau}_{ij}}. \label{EQ: shear rate}
\end{align}

\begin{figure}
    \centering
    \includegraphics[width=\textwidth]{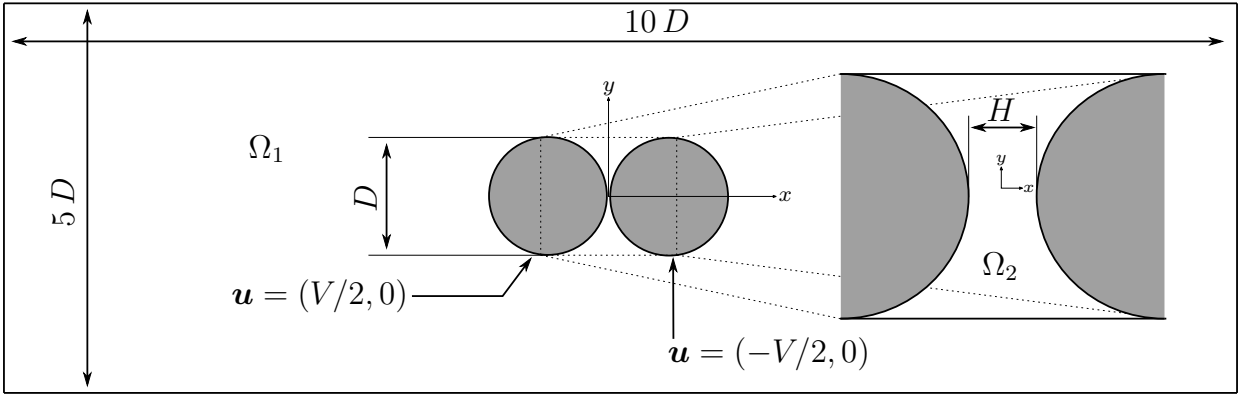}
    \caption{Schematic showing the problem geometry for the entire flow system ($\Omega_1$) investigated through
        numerical methods and the reduced system ($\Omega_2$) investigated with both analytical and numerical methods. \label{FIG:
    setup}}
\end{figure}

The problem geometries are depicted in figure \ref{FIG: setup}, where the inset highlights the portion of the
system considered in the analytical investigation. Aligning the system mid-plane in a Cartesian coordinate system
the two cylinders are placed with their centres located at $(-H/2-D/2,0)$ and $(H/2+D/2,0)$, where $H$
is the minimum separation distance and~$D$ the cylinder diameter. The computational domain for the whole system has dimensions $10\,D \times 5\,D$, which is sufficiently large for the cylinders to be essentially unconfined: waning stresses away from
the moving cylinders lead to the formation of a yield envelope in the immediate vicinity of the cylinders, outside
of which the fluid forms a rigid plug attached to the domain walls. Because the fluid in the far field is
unyielded for the range of yield stresses explored in this study, we set no-slip boundaries at $y\pm2.5\,D$ and
pressure inlets and outlets at $x\pm5\,D$. The cylinders have a constant relative approach velocity of $V$.

\subsection{Large-scale non-dimensionalisation}
\label{sec:largescalenondim}

Choosing a velocity scale of~$V$, length scale of~$D$, shear rate scale of~$V/D$, and stress scale
of~$\hat\eta V/D$, we obtain the dimensionless equations

\begin{align}
    \frac{\partial u}{\partial x} + \frac{\partial v}{\partial y} &=0, \label{EQ: continuity dimensionless} \\
    \frac{\partial \tau_{xx}}{\partial x} + \frac{\partial \tau_{xy}}{\partial y} - \frac{\partial p}{\partial x}
    &= 0,\\
    \frac{\partial \tau_{yx}}{\partial x} + \frac{\partial \tau_{yy}}{\partial y} - \frac{\partial p}{\partial y}
    &=0,
\end{align}

\begin{equation}
    \begin{cases}
        \tau_{ij} = \left(2 + \frac{\Bn}{\dot{\gamma}}\right)\dot{\gamma}_{ij} &
        \quad\mbox{if $\tau>\Bn$,} \\
        \dot{\gamma}_{ij} = 0 & 
    \quad\mbox{if $\,\tau\leq\Bn$,}
    \end{cases}
    \label{EQ: constitutive equation dimensionless}
\end{equation}
where
\begin{equation}
  \Bn := \frac{\hat\tau_Y D}{ \hat\eta V}
\end{equation}
is a Bingham number for the macroscopic flow external to the gap.

\subsection{Computational method}

We numerically compute the solution of equations~\ref{EQ: continuity dimensionless}--\ref{EQ: constitutive equation dimensionless} for two approaching cylinders with a small gap size ($H/R = 0.01$ where~$R \equiv D/2$ is the cylinder radius), and calculate the resulting forces on the cylinders.
To handle the disparate length scales of this problem in a computationally efficient manner we use the method of overset grids (also called overlapping, overlaid or Chimera grids) in a finite difference framework to discretize the domain.
This method and grid generation algorithm is discussed in detail in~\citet{Chesshire1990},~\citet{Henshaw1998}, and~\citet{Koblitz2017} where its efficacy for particulate flow simulations was demonstrated.
Briefly, the overset grid method represents a complex domain using multiple body-fitted curvilinear grids that are allowed to overlap whilst being logically rectangular.
The overlapping aspect brings flexibility and efficiency to grid generation, which is beneficial for moving body problems.
Here, since the cylinders are static, the chief benefit of the overset grid method is that the grids can be locally refined near the gap whilst keeping the grids
logically rectangular. The resultant linear systems are solved using the MUMPS library \citep{Amestoy2001}, a
massively parallel direct linear solver. We use meshes with a minimum of 15 points across the narrowest part of
the gap and cluster grid points near the cylinder surfaces and wider gap region by stretching the constituent
grids.

Applying a standard finite difference method to equations \eqref{EQ: continuity dimensionless}--\eqref{EQ: constitutive equation dimensionless} is not straightforward, due to the non-differentiable plastic dissipation term.
A straight-forward way of dealing with this numerical difficulty is \revb{to regularize} equation \eqref{EQ: constitutive equation dimensionless} by removing the singularity at~$\dot\gamma = 0$.
This approach has been used in studies of viscoplastic flows past bluff bodies,
see~\cite{Tokpavi2008,Mitsoulis2002,Mossaz2010}.
However this can yield inaccurate results, especially for lubrication-type flows or if flow stability or finite-time stoppage are of critical interest~\cite{Frigaard2005,Putz2009,Wachs2016}.
Instead, we use an iterative method based on the variational form of the Bingham problem, established by~\citet{Duvaut1972}, which forms the basis for the widely used augmented Lagrangian (AL) first proposed by \citet{Glowinski1984}.
This formulation is commonly known as ALG2 and is used extensively in the literature, see
\citet{Yu2007,Chaparian2017b,Muravleva2015} and references therein, so we do not give details here. 
For its solution we use the Uzawa type algorithm of \citet{Olshanskii2009} and \citet{Muravleva2008}.

\subsection{Lubrication flow in the gap}
\label{sec:lubrication}

The problem shown in the inset of figure~\ref{FIG: setup}, i.e.\ the narrow gap between two symmetric surfaces approaching with relative speed~$V$, has an asymptotic solution due to~\citet{Balmforth2017a}, if the gap~$H$ is small compared to the cylinder radius~$R$.
In this section we give an overview of this solution; in section~\ref{sec:results} we will compare this to fully numerical solutions both in the restricted domain (inset of figure~\ref{FIG: setup}) and the full domain.
Note that this section considers a non-dimensionalisation of the governing equations appropriate to the gap scale; the non-dimensionalisation given previously in section~\ref{sec:largescalenondim} is appropriate for the macroscopic flow.
We take~$x$ to be the coordinate across the gap and~$y$ the coordinate along the gap, consistent with the setup shown in figure~\ref{FIG: setup}.

We write~$\hat \uv \equiv (\hat u, \hat v)$ and without loss of generality
\begin{align}
    \hat \tauv  \equiv \left( \begin{array}{cc}
    \hat\sigma & \hat\psi \\
       \hat\psi     & -\hat\sigma
       \end{array} \right).
\end{align}
Following the approach in~\cite{Balmforth2017a}, variables are scaled as
\begin{align}
    x = \hat{x}/\mathcal{H},\quad
    y = \hat{y}/\mathcal{L},\quad
    u = \hat{u}/\mathcal{U},\quad
    v = \hat{v}/(\mathcal{U}/\epsilon),\quad
    p = \hat{p}/\mathcal{P},
\end{align}
where $\epsilon\equiv\mathcal{H}/\mathcal{L}$ is a small parameter.
This implies the scaled continuity equation is
\begin{equation}
    \frac{\partial u}{\partial x} + \frac{\partial v }{\partial y} = 0.
\end{equation}
The stress scale is chosen as $\tau = \hat\tau / (\epsilon \mathcal{P})$, which implies
\begin{align}
    \frac{\partial p}{\partial x} = \epsilon\frac{\partial \sigma}{\partial x} + \epsilon^2\frac{\partial
\psi}{\partial y}, \quad
    \frac{\partial p}{\partial y} = \frac{\partial \psi}{\partial x} + \epsilon\frac{\partial\sigma}{\partial y},
    \label{EQ: lub force balance}
\end{align}
so that the main force balance (to~$O(\epsilon)$) is between the axial pressure gradient and transverse shear stress   gradient.
Strain rates are scaled by~$(\mathcal{U}/\epsilon)/\mathcal{H}$, giving
\begin{equation}
    \gdot=\sqrt{\frac{1}{4}\left(\epsilon^2\frac{\partial u}{\partial y}+\frac{\partial v}{\partial
    x}\right)^2+\epsilon^2\left(\frac{\partial u}{\partial x}\right)^2}.
\end{equation}
The above scaling implies in the yielded regions
\begin{equation}
    \tau_{ij} = \left( \frac{2\hat\eta \mathcal{U}}{\epsilon^2 \mathcal{P} \mathcal{H}} + \frac{\hat\tau_Y}{\epsilon
    \mathcal{P} \gdot} \right) \dot{\gamma}_{ij}.
\end{equation}
The velocity scale is set by the motion of the cylinders as $\mathcal{U} := V$ 
and therefore the pressure scale is chosen as
\begin{equation}
  \mathcal{P} := \frac{\hat\eta V}{\epsilon^2 \mathcal{H}}.
\end{equation}
We additionally fix the characteristic length and gap scales as $\mathcal{L}=R$ and $\mathcal{H}=H$,
respectively.
This gives the scaled constitutive equation as
\begin{equation}
  \label{EQ: scaledconstitutive}
    \begin{cases}
        \tau_{ij} = \left(2 + \frac{\Bnlub}{\gdot}\right)\dot{\gamma}_{ij} &
        \quad\mbox{if $\tau>\Bnlub$,} \\
        \dot{\gamma}_{ij} = 0 & 
    \quad\mbox{if $\,\tau\leq\Bnlub$,}
    \end{cases}
\end{equation}
where
\begin{equation}
  \Bnlub := \frac{\hat\tau_Y}{\epsilon \mathcal{P}}
\end{equation}
is a Bingham number for the squeeze flow in the gap.
Note that~$\Bnlub / \Bn = \epsilon^2/2$; the squeeze flow `sees' a much lower Bingham number than the macroscopic flow around the cylinders.

\subsubsection{Leading-order solution}
The components of the shear rate tensor are
\begin{align}
  \psi &\equiv \gdot_{xy} = \frac{1}{2}\left(\epsilon^2\frac{\partial u}{\partial y} + \frac{\partial v}{\partial x} \right),\\
  \sigma &\equiv \gdot_{xx} = \epsilon \frac{\partial u}{\partial x}.
\end{align}
Therefore, discarding terms of~$O(\epsilon)$, the shear rate magnitude is
\begin{equation}
  \gdot = \frac{1}{2}\left| \frac{\partial v}{\partial x} \right|,
\end{equation}
and in the fully yielded part of the flow $\psi \gg \sigma$.
Equation~\ref{EQ: scaledconstitutive} is used to write
\begin{equation}
    \psi=\frac{\partial v}{\partial x} + \Bnlub\sign\left(\frac{\partial v}{\partial x}\right),
\end{equation}
and the main force balance reduces to
\begin{align}
  \frac{\partial p}{\partial x} = 0 \Rightarrow p=p(y),\quad
  \frac{\partial p}{\partial y} = \frac{\partial \psi}{\partial x} \Rightarrow \psi=x\frac{\partial p}{\partial y},
\end{align}
the constant vanishing by symmetry, meaning that the pressure gradient is, to leading order, constant across the gap and balanced along the gap
by the transverse shear stress.
Exploiting the symmetry of the configuration, in the quadrant $x>0$, $y>0$ we must then have $v>0$,
$\frac{\partial v}{\partial x}<0$, and so from the main force balance and constitutive law we find the velocity
profile across the gap
\begin{equation}
    \frac{\partial v}{\partial x} = x\frac{\partial p}{\partial y} + \Bnlub,
\end{equation}
which may be integrated to give
\begin{equation}
v = \begin{cases}
    -\frac{1}{2}\frac{\partial p}{\partial y} \left(\frac{1}{h} - x\right)\left(\frac{1}{2}h - 2X + x\right), \ X
    < x \leq \frac{1}{2}h(y) \\
    -\frac{1}{2}\frac{\partial p}{\partial y}\left(\frac{1}{2}h - X\right)^2, \ 0 \leq x \leq X,
\end{cases}
\end{equation}
where~$X \equiv \Bnlub/|\frac{\partial p}{\partial y}|$ is the plug boundary location, and we have used a no-slip boundary
condition at the cylinder surface, located at~$x = \frac{1}{2}h(y)$.
The continuity equation and boundary conditions imply a flow rate constraint
\begin{equation}
\frac{\partial}{\partial y} \int_{-\frac{1}{2}h}^{\frac{1}{2}h} v\ \text{d}x = 1
\end{equation}
which, when evaluated using the velocity solution, gives a cubic equation for the pressure
gradient~$\frac{\partial p}{\partial y}(y)$:
\begin{equation}
    -\frac{1}{12}\frac{\partial p}{\partial y}(h + X)(h - 2X)^2 = y.
\label{eq:lubrication:px}
\end{equation}

It can be shown that the plug in the region~$|x|<X$ undergoes~$O(\epsilon)$ plastic flow, which is not present in the  above asymptotic solution.
This may be recovered by keeping terms~$O(\epsilon)$ and is sometimes referred to as a pseudo-plug; it does not change the equation for the pressure gradient to leading order~\citep{Balmforth2017a}.

For two converging cylinders the non-dimensional separation distance is
\begin{equation}
    h(y) = 1 + \frac{2}{\epsilon}\left(1 - \sqrt{1 - y^2}\right), \ 0 \leq |y| < 1. \label{EQ: separation}
\end{equation}
We numerically evaluate equation~\ref{eq:lubrication:px} to compute~$\frac{\partial p}{\partial y}(y)$ and
thence~$p(y)$, with an additional ambient pressure constraint outside the disks enforced as~$p(1) = 0$.
The leading-order lubrication force is then numerically computed as~$2\int_0^1 p\,\mathrm{d}y$.

\subsection{Flow field diagnostics}
In order to classify the structure of the numerically-calculated flowfields we make use of an invariant measure of the velocity gradient
tensor that gives an indication of the relative strength of the shear rate tensor and vorticity field \citep{Davidson2004}
\begin{equation}
    Q=-\frac{1}{2}\frac{\partial \hat{u}_i}{\partial \hat{x}_j}\frac{\partial \hat{u}_j}{\partial \hat{x}_i} =
    -\frac{1}{2}\left(\hat{\dot{\gamma}}_{ij}\hat{\dot{\gamma}}_{ij}-\frac{1}{2}\hat{\boldsymbol{\omega}}^2\right), \label{EQ: flow topology}
\end{equation}
where $\hat{\boldsymbol{\omega}}$ is the vorticity. We use the normalized form of \eqref{EQ: flow topology}
\begin{equation}
    \Lambda = \frac{\hat{\dot{\gamma}}_{ij}\hat{\dot{\gamma}}_{ij} -
    1/2(\hat{\boldsymbol{\omega}}^2)}{\hat{\dot{\gamma}}_{ij}\hat{\dot{\gamma}}_{ij}+1/2(\hat{\boldsymbol{\omega}}^2)},
\end{equation}
such that values of $\Lambda=-1,0,1$ correspond to flow dominated by rotation, shear, and strain, respectively
\citep{De2017}. 

The rate of working the fluid, $\hat{\dot{W}}$, is calculated by integrating the rate of viscous dissipation,
$\hat{\Phi}=\hat{\tau}_{ij}\hat{\dot \gamma}_{ij}$,
over a suitable control volume
\begin{equation}
    \hat{\dot{W}}(\Omega) = \int_{\Omega-V_C}\hat{\tau}_{ij}\hat{\dot{\gamma}}_{ij}\,\mathrm{d}V,
\end{equation}
where $V_C$ is the volume occupied by the cylinders. This is scaled by the force on the cylinders and the closing
velocity, $\mathcal{W}= FV$, while the viscous
dissipation is scaled using a characteristic energy density scale
$\mathcal{E}=\eta {V}^2/\mathcal{H}^2$.


\section{Results}
\label{sec:results}

We investigate the squeeze flow between two infinite circular cylinders in three different cases based on the set-up shown in figure~\ref{FIG: setup}.
Non-dimensionalisation is as described in section~\ref{sec:largescalenondim}.
The external Bingham number $\Bn$ is varied between 0 and 2000 in all cases, 
with the minimum separation distance kept
constant at $0.01$ non-dimensional units (i.e.\ 1\% of the cylinder radius), resulting in a gap Bingham number
$\Bnlub$ range of 0 to 0.1.

Two cases use the full computational domain, labelled as $\Omega_1$ in figure \ref{FIG: setup}, with differing far
field boundary conditions. The quiescent case (meaning here that the flow is zero outside a finite yield envelope) has velocity outlets at the vertical domain
boundaries, imposing an ambient pressure of $p=0$. No-slip and no-penetration conditions are imposed on the
horizontal domain boundaries, allowing the cylinders to be surrounded by a bounded yielded region enclosed by a yield envelope.

The shear flow case considers the same geometry as the quiescent case but with the introduction of a
macroscopic flow to raise the stress above the material yield stress in the far field, thus removing the yield
envelope. This is done by imposing wall velocities $\pm U_w$ on the horizontal domain boundaries, resulting in a
macroscopic shear rate of $\dot{\gamma}=5$.

Finally, we consider the reduced domain labelled as $\Omega_2$ in figure \ref{FIG: setup}, only including the gap
between the cylinders. No-slip and no-penetration conditions are applied on the cylinder surfaces, and symmetry conditions at $x=\pm 1$ in a similar manner to \citet{Frigaard2004} and \citet{Muravleva2015}.

We begin in section~\ref{subsec: flow field kinematics} with a detailed description of the flow field kinematics for the two cylinder system using the 
DNS results in the full computational domain. We then investigate the validity of the small gap approximation 
used to develop a leading order viscoplastic lubrication solution in section~\ref{subsec: small gap
approximation}. Following this we compare
pressure profiles along the the axis of symmetry, and the resultant normal force exerted on the
cylinders, to solutions from viscoplastic lubrication theory in section~\ref{subsec: pressure profiles}. Finally, we investigate viscous dissipation in the system in section~\ref{subsec: viscous dissipation}.


\subsection{Flow field kinematics}\label{subsec: flow field kinematics}

\begin{figure}
\centering
\includegraphics[width=\textwidth]{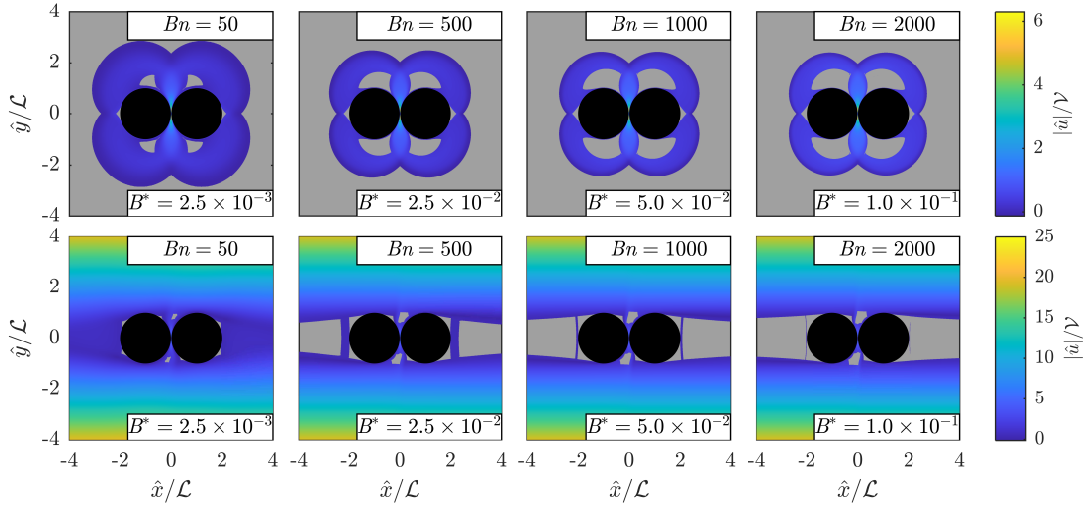}
\caption{Velocity magnitude plots for increasing $\Bn$ (left to right) with quiescent conditions (top row) and
macroscopic shear rate $\dot{\gamma}=5$ (bottom row). \label{FIG: yield surfaces}}
\end{figure}

\begin{figure}
    \centering
    \includegraphics[width=\textwidth]{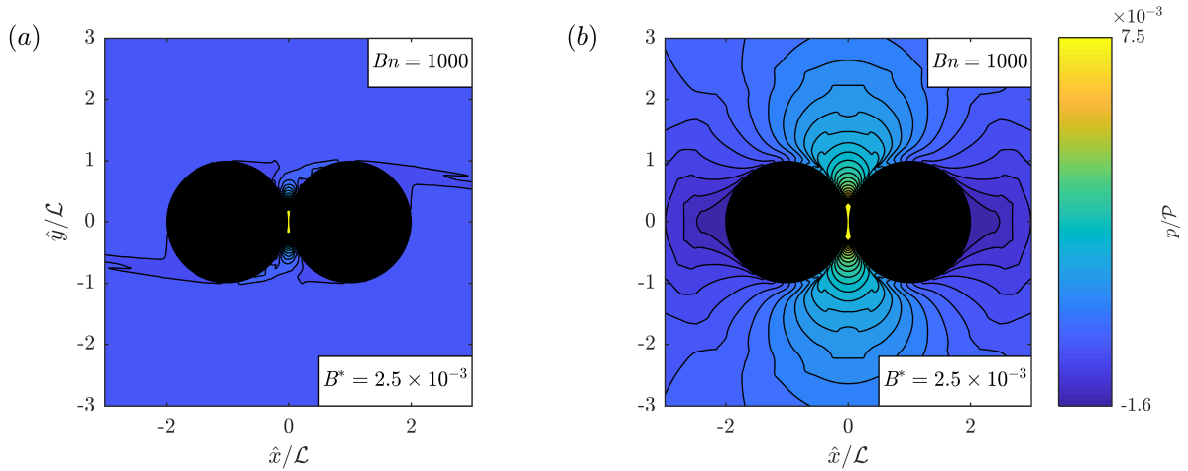}
    \caption{Pressure contours for the two cylinder system in macroscopic shear flow (left) and quiescent (right) with $\Bn=1000$. \label{FIG: pressure contours}}
\end{figure}

Figure \ref{FIG: yield surfaces} shows a binary yielded/unyielded mask in grey overlaid on colour maps of the
velocity magnitude for the quiescent and sheared systems, with the Bingham number increasing from left
to right. The unyielded regions are identified as areas where the second invariant of the shear stress falls below
the yield stress, plus some small constant which we take as $0.1\,\%$ of the yield stress.

The top row in figure~\ref{FIG: yield surfaces} corresponds to the quiescent case, where for $\Bn\geq50$ classical features
of moving bodies in yield stress fluids can be seen: unyielded caps on the stagnation points, unyielded plugs in
the equatorial planes of the cylinders, and a yield envelope fully surrounding the two cylinder system
\citep{Tokpavi2008,Putz2010,Chaparian2017b,Beris1985,Ansley1967,Adachi1973}. As the Bingham number increases the
unyielded stagnation caps and the equatorial plugs grow while the yield envelope shrinks.

The bottom row of figure~\ref{FIG: yield surfaces} corresponds to the shear flow case, where the background shear flow has noticeably changed the yield surface
features: stagnation points have shifted, leading to two caps on the rear of the
cylinders, placed symmetrically about the longitudinal axis, and one on the front of each cylinder.
The equatorial plugs are no longer present, but two unattached plugs have formed in the gap openings, placed asymmetrically about
the longitudinal axis.
For $\Bn>50$ central plugs can be seen fore and aft of the two cylinder system.

Figure \ref{FIG: pressure contours} show contour plots of the pressure field for a quiescent (right panel) 
and shear flow (left panel) case at $\Bn=1000$, with the contours drawn at the same levels in both panels.
The quiescent case shows a pressure drop from the gap to the rear stagnation cap, with roughly equally 
spaced iso-contours along shear layers attached to the cylinder surfaces and along the yield envelope boundary. 
In contrast, the shear flow case shows a rapid pressure decay along the gap with a more uniform pressure 
field outside of the gap.

\subsection{Small gap approximation} \label{subsec: small gap approximation}

In section~\ref{sec:lubrication} a lubrication approximation was constructed which has, to leading order, pressure constant across the gap. This approximation relies on the gap being small, 
viz.~$\epsilon\ll1$, which is satisfied at the gap centre. However, since the approaching 
surfaces are elliptic (see equation \eqref{EQ: separation}), this condition will be violated towards
the gap exit.

\begin{figure}
    \centering
    \includegraphics[scale=1]{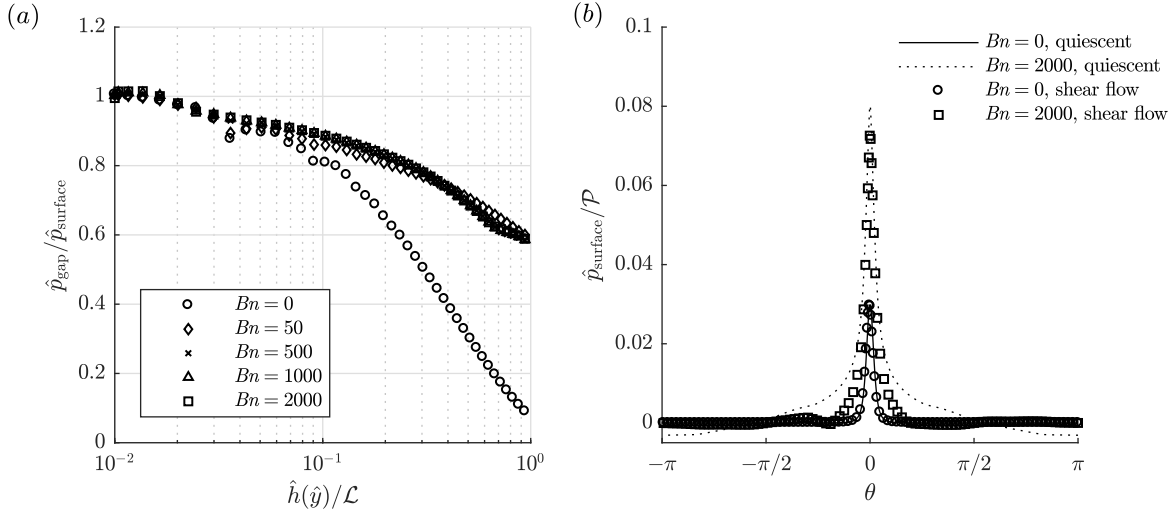}
    \caption{Left: Ratio of centreline pressure to surface pressure as a function of the local gap width for
        $\Bn=0$ (circles), $50$ (diamonds), $500$ (crosses), $1000$ (triangles), and $2000$ (squares). 
        Right: Pressure distributions over the entire cylinder surface for the limiting $\Bn=0$ and $\Bn=2000$
        quiescent flow cases as a function of angle away from the gap centre.
        \label{FIG: pressure ratio}}
\end{figure}

We investigate the validity of this constant pressure solution by examining the ratio of the pressure at the
gap centre to that at the surface of the cylinder as a function of the local gap width. 
In the left panel of figure \ref{FIG: pressure ratio} we plot this gap-to-surface pressure ratio against the 
normalized gap width as a function of $y$ as markers for
$\Bn=0$--$2000$ ($\Bnlub=0$--$0.01$). 


For all cases, the gap-to-surface pressure ratio slowly decreases towards the gap exit, but is above $0.8$ until approximately
$\hat{h}(\hat{y})/\mathcal{L} > 0.1$. In the Newtonian case, the pressure ratio then rapidly decreases
away from the gap centre, becoming
negligible at the gap exit. All viscoplastic cases show
similar behaviour to one another (as indicated by the marker overlap in the left panel of 
figure~\ref{FIG: pressure ratio}). The gap-to-surface pressure ratio remains 
close to unity close to the gap centre before
slowly decreasing. Unlike for the Newtonian case, no rapid decrease is found when 
$\hat{h}(\hat{y})/\mathcal{L} > 0.1$ and as a result the gap-to-surface pressure remains above $0.8$ further 
along the gap for the viscoplastic cases.


The right panel of figure~\ref{FIG: pressure ratio} shows the pressure distributions over the entire cylinder
surface, with the gap centre located at $\theta = 0$ and the gap exits at 
$\theta = (-\pi/2,\pi/2)$, for the limiting quiescent and shear flow cases.
The Newtonian surface pressure distributions of the quiescent and shear flow cases overlap, showing a peak 
at the gap centre and a rapid decay towards the exits. For the high yield stress cases, the pressure 
distributions of the quiescent and shear flow cases are broadly similar, both showing a pressure peak in the gap
centre. However, towards the gap 
exit the pressure decays more slowly for the quiescent case than for the shear flow case. 

In isolation, the left panel of figure~\ref{FIG: pressure ratio} shows that the leading-order solution with constant pressure across the gap, presented in 
section~\ref{sec:lubrication}, is valid for only a small portion of the gap between the approaching cylinders,
particularly in the absence of a yield stress. However, from the surface pressure distributions 
in the right panel of figure~\ref{FIG: pressure ratio} it is clear that the overwhelming contribution to the lubrication force comes from a narrow band in
the gap, where the gap-to-surface pressure ratio is above $0.9$ for all cases. Therefore we expect the 
leading-order solution to capture the lubrication force to a good approximation.


\FloatBarrier
\subsection{Pressure profiles in the gap} \label{subsec: pressure profiles}

\begin{figure}
\centering
\includegraphics[width=\textwidth]{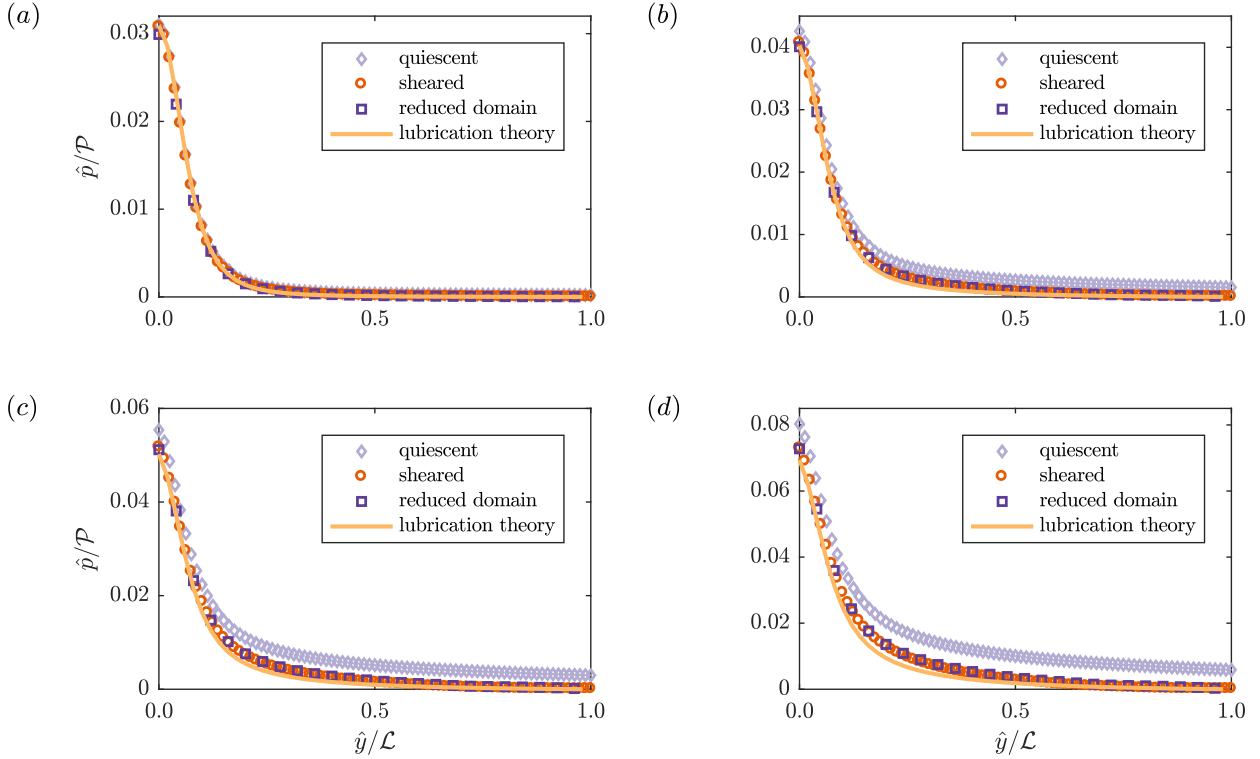}
\caption{Pressure profile along the gap centreline for quiescent (diamonds), sheared (circles), and reduced
domain (square) systems with viscoplastic lubrication solution overlayed (solid line) with $\Bn=50,500,1000,2000$
in plots ($a$) through ($d$), respectively.
}
\label{FIG: pressure profiles}
\end{figure}

Figure \ref{FIG: pressure profiles} presents pressure profiles through the centre of the gap, i.e. along the
axis of symmetry. The direct numerical
simulation (DNS) in the reduced domain gives a pressure profile in excellent agreement with the DNS of the full system in the macroscopic shear flow.
Both these cases are in good agreement with the asymptotic solution from lubrication theory: peak pressures in the centre of the gap match well for the full $\Bn$ range explored.
At higher $\Bn$, the DNS pressures of the shear flow and reduced domain cases remain in agreement but decay more slowly than the asymptotic solution as the gap widens up; this is where the lubrication approximation no longer holds.

The pressure profiles for the DNS of the full system in the quiescent case are markedly different to the asymptotic solution for $\Bn>50$: 
higher peak pressures and slower pressure decay are evident, as is an exit pressure significantly higher than the ambient pressure (which is $0$).

The relative change in peak pressure and pressure decay for increasing $\Bn$
discussed above is evident in the surface pressure distribution shown in figure~\ref{FIG: pressure ratio}. 
Moreover, it is evident that the pressure
contribution outside of the nominal gap region is negligible. 
Note that for~$\Bn=2000$ the pressure profiles look somewhat similar in magnitude between the quiescent and sheared cases.
In fact their integrals differ by about a factor of two, implying a factor of two difference in the repulsive force; this is discussed next.

\begin{figure}
\centering
\includegraphics[width=\textwidth]{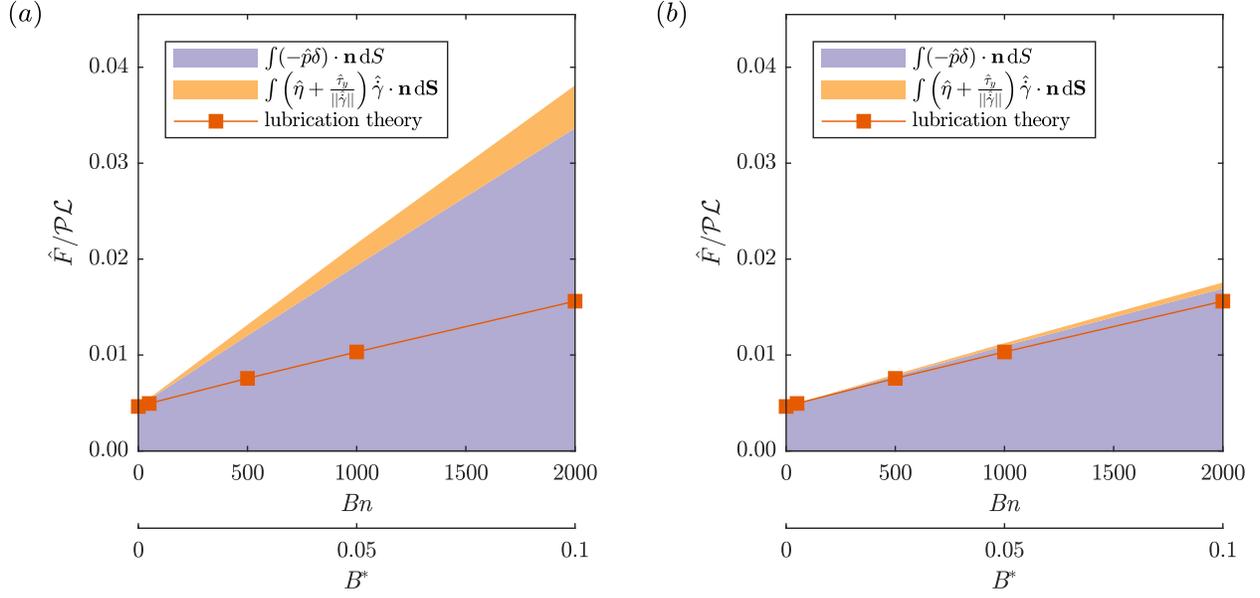}
\caption{Stacked area plots of the total drag force on a cylinder as a function of~$\mathit{Bn}$ for $(a)$ quiescent
and $(b)$ shear flow background conditions. Overlaid are the predictions from viscoplastic lubrication theory (squares). \label{FIG: drag force plots}}
\end{figure}

Figure \ref{FIG: drag force plots} presents stacked area plots of the total drag force exerted on a cylinder, decomposed into pressure and viscous contributions.
From the left panel it is clear that the force on the cylinders in quiescent fluid is dramatically underestimated when the lubrication flow in the gap is considered in isolation.
Both viscous and pressure contributions increase with $\Bn$, but the viscous contribution remains small compared to that of the pressure.
Discounting the viscous friction, the pressure alone---which remains localised to the gap---causes a more than two-fold increase in the drag force over the predictions from lubrication theory.
However when a macroscopic shear flow is added (the right panel in figure~\ref{FIG: drag force plots}), the total drag force is close to the asymptotic solution.
This mirrors the trend found in the pressure profiles in figure~\ref{FIG: pressure profiles}.

\begin{figure}
    \centering
    \includegraphics[width=.5\textwidth]{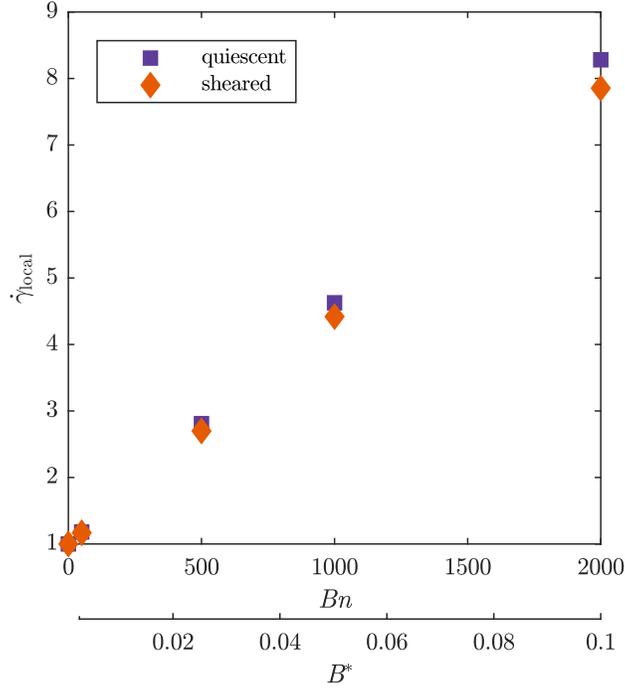}
    \caption{Local shear rates in the gap centre for quiescent (squares) and sheared (diamonds) systems for
    $\Bn=0,50,500,1000,2000$.\label{FIG: local shear}}
\end{figure}

Figure \ref{FIG: local shear} shows local shear rates, $\dot{\gamma}_{\text{local}}$, for both the sheared and
quiescent cases for the full range of $\Bn$. We define the local shear rate as an average over a $H\times 2H$ area in the
centre of the gap. The dramatic increase in pressure and drag force is not reflected in the local shear rate: the
local shear rates in the quiescent and sheared cases remain in close agreement throughout the $\Bn$ range
explored. From this we can conclude that the macroscopic flow does not affect the velocity field in the gap.
The above computations were also performed with a macroscopic shear rate one order of magnitude higher, showing
no appreciable differences in the pressure and force results discussed above. 

\FloatBarrier
\subsection{Viscous dissipation} \label{subsec: viscous dissipation}

\begin{figure}
    \centering
    \includegraphics[width=\textwidth]{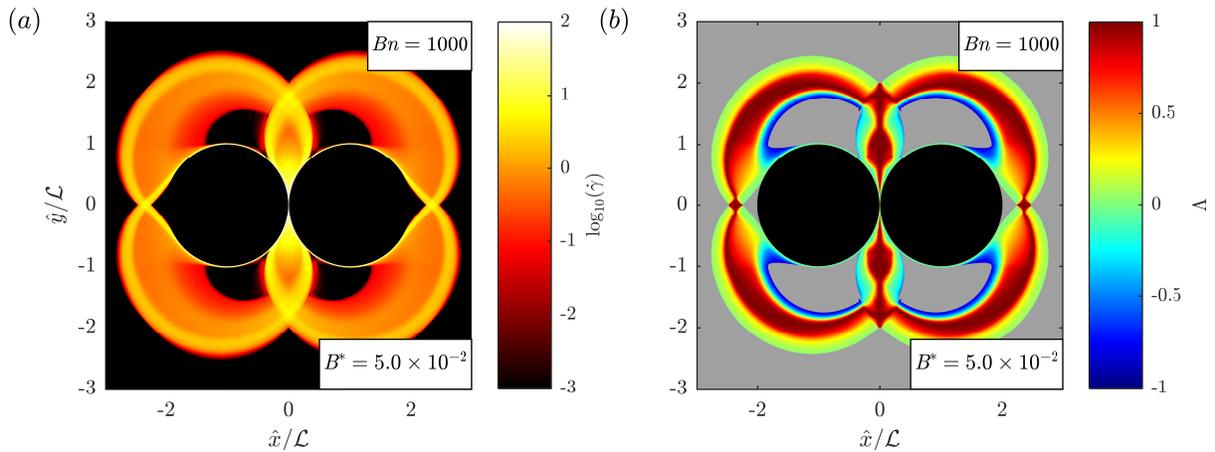}
    \caption{(a) Colour map of $\log_{10}(\dot{\gamma})$ and (b) normalized second
    invariants of the velocity gradient tensor with unyielded areas masked in grey with $\Bn=1000$ (quiescent case).\label{FIG: dissipation and si}}
\end{figure}


\begin{figure}
    \centering
    \begin{minipage}[t]{.49\textwidth}
        \includegraphics[width=\textwidth]{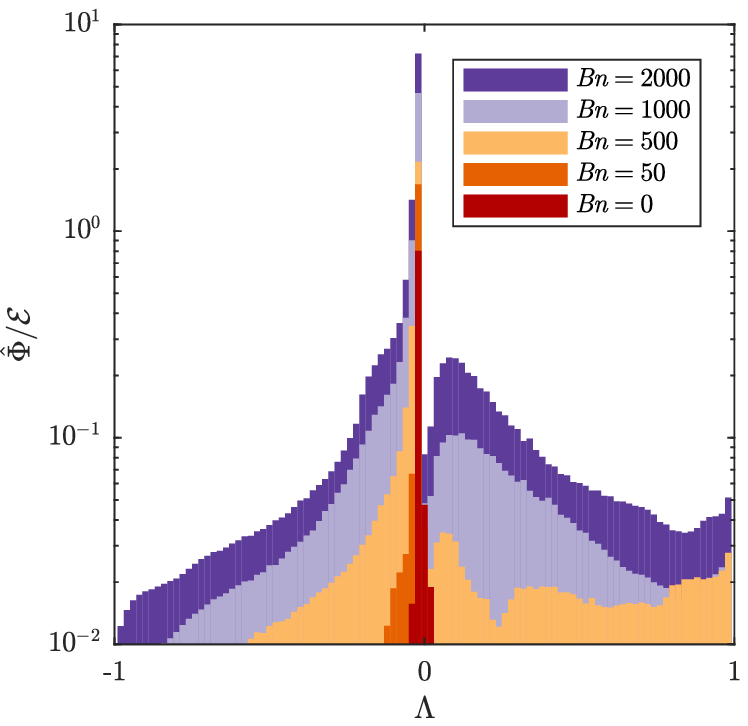}
    \end{minipage}%
    \hfill
    \begin{minipage}[t]{.49\textwidth}
        \includegraphics[width=\textwidth]{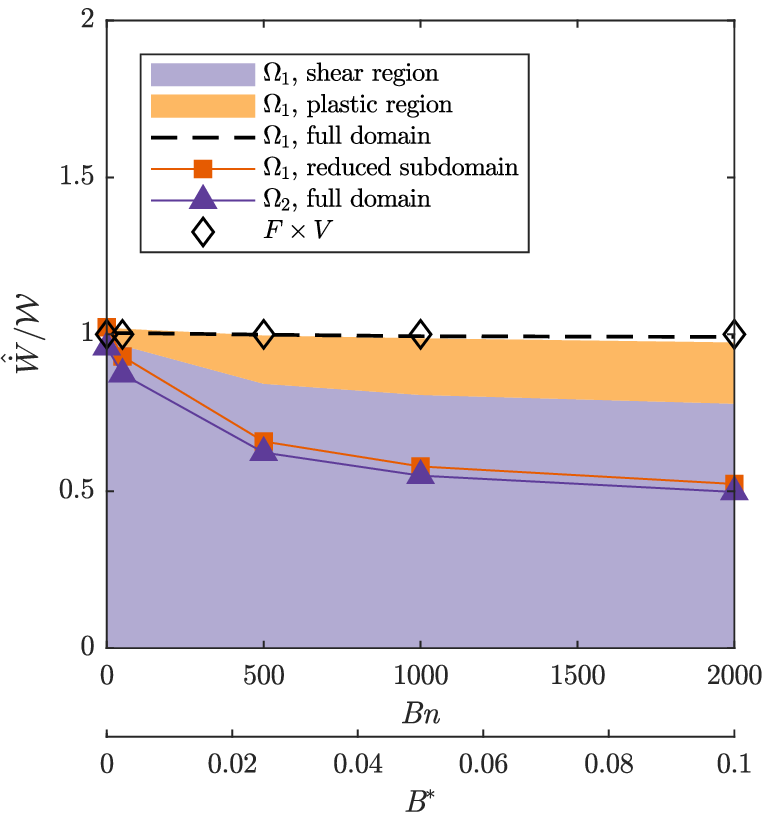}
    \end{minipage}
    \caption{Left: rate of mechanical dissipation per unit $\Lambda$. Right: stacked area plot showing the rate of work done by viscous dissipation in the shear and plastic regions, scaled by~$\mathcal{W}=FV$, with markers
        indicating the power required to move the cylinders with the approach velocity (diamonds), the total
        viscous dissipation in the system (dashed line), the viscous dissipation in the gap region of the full
        system (squares) and the total viscous dissipation in the reduced system (triangles).\label{FIG:
    energy}}
\end{figure}


The left panel of \ref{FIG: dissipation and si} shows a colour map of $\log_{10}(\gdot)$ in a quiescent case, while the right panel 
shows a contour plot of the normalized second invariant
of the velocity gradient tensor, indicating where the fluid is irrotational (red), rotational (blue) and being
sheared (green), with the yield surface overlaid.
As expected, the plugs either side of the cylinders undergo rigid body rotation. Strain rates are highest in the
thin shear layers along the cylinder surface, along the yield envelope wall, and surrounding the jet of fluid
squeezed out of the gap. Strain dominated regions are
found in the core of the fluid jet squeezed out of the gap, and in the regions between the rotating plugs and the
yield envelope. The strain rate in these plastic flow regions is orders of magnitude lower than in the adjacent
shear layers.

We turn now to the energy dissipation in the fluid.
The left panel of figure \ref{FIG: energy} shows the rate of mechanical dissipation as a function of the topology parameter
$\Lambda$. As would be expected, no dissipation occurs in the regions undergoing rigid body rotation 
($\Lambda = -1$). Some dissipation is evident in the irrotational regions ($\Lambda = 1$) at high 
Bingham numbers, this is attributed to pseudo-plug regions where the flow is held close to the 
yield stress \citep{Walton1991}.

In all cases the dissipation is 
highest in regions of shear, peaking at $\Lambda = 0$. While the rate of mechanical dissipation decreases monotonically as the flow becomes rotationally dominated, for $\Bn>0$ as the flow becomes dominated by strain there exists a second, small peak in dissipation which decays slowly as $\Lambda$ approaches 1.

The right panel of figure \ref{FIG: energy} shows the rate of work done on the fluid in different regions and flow
structures for the full and reduced systems, as well as the power required to move the cylinders with the set
approach velocity. Shear regions and plastic regions have been defined as areas where $-1/3\leq\Lambda\leq1/3$ and
$\Lambda>1/3$, respectively \citep{De2017}. While the dissipation in plastic flow is small compared
to that in shear flow, it still forms a significant source of viscous dissipation outside of the gap area due to the size of the plastic flow regions (see figure \ref{FIG: dissipation and si}).
Finally, the rate of work done in the gap region is very similar for both the full ($\Omega_1$) and reduced ($\Omega_2$) domains.
This shows that the excess drag force on the cylinders in the full domain compared to lubrication theory (figure~\ref{FIG: drag force plots}$(a)$) arises from energy dissipation external to the gap.

\section{Conclusions}
\label{sec:conclusions}

In this paper we have presented results on the squeeze flow between two infinite circular cylinders in a Bingham fluid, which we use as a simple model for the flow of non-colloidal particles in a viscoplastic fluid.
Understanding this flow is essential to building models of the large-scale flow of such suspensions.
Although the calculations presented here have been two-dimensional, we expect similar phenomena will occur in three dimensions (where the particles would be spheres).

In section~\ref{sec:results} we presented results from three numerical experiments: two modelling the approaching
cylinders within a quiescent and a sheared fluid, and one modelling just the gap between the approaching cylinders,
removing any external influence. We showed that unlike for a Newtonian fluid, the macroscopic flow external to the gap has a large effect on the lubrication forces felt by two cylinders in near-contact.
In a quiescent Bingham fluid, the lubrication forces were approximately double those predicted by viscoplastic
lubrication theory, but were still caused primarily by the localised high lubrication pressure in the gap, as for a Newtonian fluid.
The high lubrication pressure compared to theory is due to the enclosing yield envelope which forms around the two particle system and
causes a recirculating flow, introducing significant viscous dissipation into the system. Most of the extra viscous
dissipation occurs in shear layers along the cylinder surface and yield envelope walls, although at high Bingham
numbers the contribution from plastic flow regions near the yield envelope becomes appreciable. 

Introducing a macroscopic shear flow or modelling just the gap area between the cylinders gave nearly identical
results, and agreed closely with the predictions from lubrication theory.
We conclude that the background shear flow acts to eliminate the yield envelope in the macroscopic flow around the
particles.
This in turn removes the recirculating flow and complex flow structures, \revb{where} large sources of viscous
dissipation in the quiescent case \revb{appear}, and hence lowers the lubrication pressure and resulting lubrication force.
The resulting pressure profiles in the gap are well-described by lubrication theory local to the gap. The
results indicate that the macroscopic shear rate does not appreciably affect the velocity field in the narrow gap
region. This
conclusion is insensitive to the exact macroscopic shear rate used, provided the yield envelope is removed.

The above implies that lubrication force models using an effective viscosity based on the
local shear rate (such as the approach used for shear-thinning fluids in~\citet{Vazquez-Quesada2016})
may not be accurate for viscoplastic fluids.
Instead, we suggest the use of sub-grid-scale lubrication force models based on viscoplastic lubrication theory,
with the understanding that they may become invalid in regions without a macroscopic stress above the yield stress, i.e.\ where particles become confined by their own yield envelopes.
This will allow for a large range of validity, for example in simulations of the type considered in~\cite{Vazquez-Quesada2016,Bian2014a} among others, where a dense suspension is subject to shear, and a sub-grid-scale lubrication force model is needed due to the close particle-particle approaches.
However in other cases, for example dilute particulate suspensions sedimenting in a quiescent fluid, we have shown in this paper that a sub-grid-scale lubrication force model based solely on lubrication theory in the gap may not be appropriate.
Until a more sophisticated sub-grid-scale model is developed, the only current option is DNS computations with sufficiently high resolution in the inter-particle gaps.

\appendix 

\FloatBarrier
\bibliography{interacting_disks}

\end{document}